\documentclass[cameraready]{Interspeech}

\title{Calibration-Reasoning Framework for Descriptive Speech Quality Assessment}

\author[affiliation={1,2}]{Elizaveta}{Kostenok}
\author[affiliation={1}, orcid=0000-0002-8347-8637]{Mathieu}{Salzmann}
\author[affiliation={2}, orcid=0000-0002-5569-9491]{Milos}{Cernak}

\address{
    $^1$ EPFL, Lausanne, Switzerland \\
    $^2$ Logitech, Lausanne, Switzerland
}

\email{lisakostenok@gmail.com, mathieu.salzmann@epfl.ch, mcernak@logitech.com}


\newcommand{\blue}[1]{\textcolor{blue}{#1}}

\usepackage{comment}

\usepackage{comment}
\usepackage{booktabs}
\usepackage{multirow}
\usepackage{amssymb}
\usepackage{amsfonts}
\usepackage{amsmath}
\usepackage{bbm}
\usepackage{balance}
\begin{document}

\maketitle

\begin{abstract}
Explainable speech quality assessment requires moving beyond Mean Opinion Scores (MOS) to analyze underlying perceptual dimensions. To address this, we introduce a novel post-training method that tailors the foundational Audio Large Language Model for multidimensional reasoning, detection and classification of audio artifacts. First, a calibration stage aligns the model to predict predefined perceptual dimensions. Second, a reinforcement learning stage leverages Group Relative Policy Optimization (GRPO) with dimension-specific rewards to heavily enhance accuracy of descriptions and temporal localization of quality issues. With this approach we reach state-of-the-art results of \textbf{0.71} mean PCC score on the multidimensional QualiSpeech benchmark and \textbf{13\%} improvement in MOS prediction driven by RL-based reasoning. Furthermore, our fine-grained GRPO rewards substantially advance the model's ability to pinpoint and classify audio artifacts in time.
\end{abstract}

\section{Introduction}
Non-intrusive speech quality assessment methods mainly target MOS score, as closest measure of human perception. While deep learning models  approximate it with high accuracy~\cite{9414878, saeki2022utmos,ragano2024scoreq}, their "black-box" predictions lack interpretability. Efforts to bridge this gap decompose quality into perceptual dimensions like intelligibility and naturalness \cite{mittag2021nisqa, 9746108, 10096680, wardah2025sqast}. Although these approaches offer a more nuanced perspective, they are restricted to quantitative assessment; they can neither characterize the specific types of audio artifacts nor localize them within the audio recording.


Audio Large Language Models (Audio LLMs) have recently shifted the field toward Explainable MOS -- generating descriptive assessments to interpret these scores~\cite{wang-etal-2025-qualispeech, monjur2025speechqualityllmllmbasedmultimodalassessment, chen2025audio, wang2025speechllmasjudgesgeneralinterpretablespeech}. Beyond outputting verifiable numerical values, these models can leverage natural language reasoning to classify artifacts and localize them in time. However, current Explainable MOS systems prioritize conversational fluency over diagnostic precision. Because the descriptive quality assessment task is absent from their pre-training mixtures, their reasoning is often ungrounded; their MOS prediction accuracy typically falls short of traditional score-based methods due to interference from hallucinated or incorrect dimension-wise predictions. To serve as a reliable diagnostic tool rather than a general audio quality summarization, these models require targeted alignment that strictly enforces both dimensional and temporal accuracy.

To bridge this gap, we introduce a two-stage post-training methodology comprising a Calibration stage to learn dimension-wise scales, and a Reasoning stage implemented via GRPO \cite{shao2024deepseekmath} with fine-grained, dimension-specific rewards. When applied to the Audio Flamingo 3 model \cite{ghosh2025audio}, Calibration-Reasoning framework\footnote{Audio samples and generated assessments are available at \url{https://calibrationreasoningframework.github.io/audiollmdemo/}; model weights are hosted at \url{https://huggingface.co/kostl/af3-sqa-grpo}, with source code to be released upon publication.} sets a new state-of-the-art on QualiSpeech \cite{wang-etal-2025-qualispeech} benchmark, outperforming existing baselines in dimension-wise judgments and the temporal localization of audio artifacts.

\begin{figure*}[t]
    \centering
    \includegraphics[width=\textwidth]{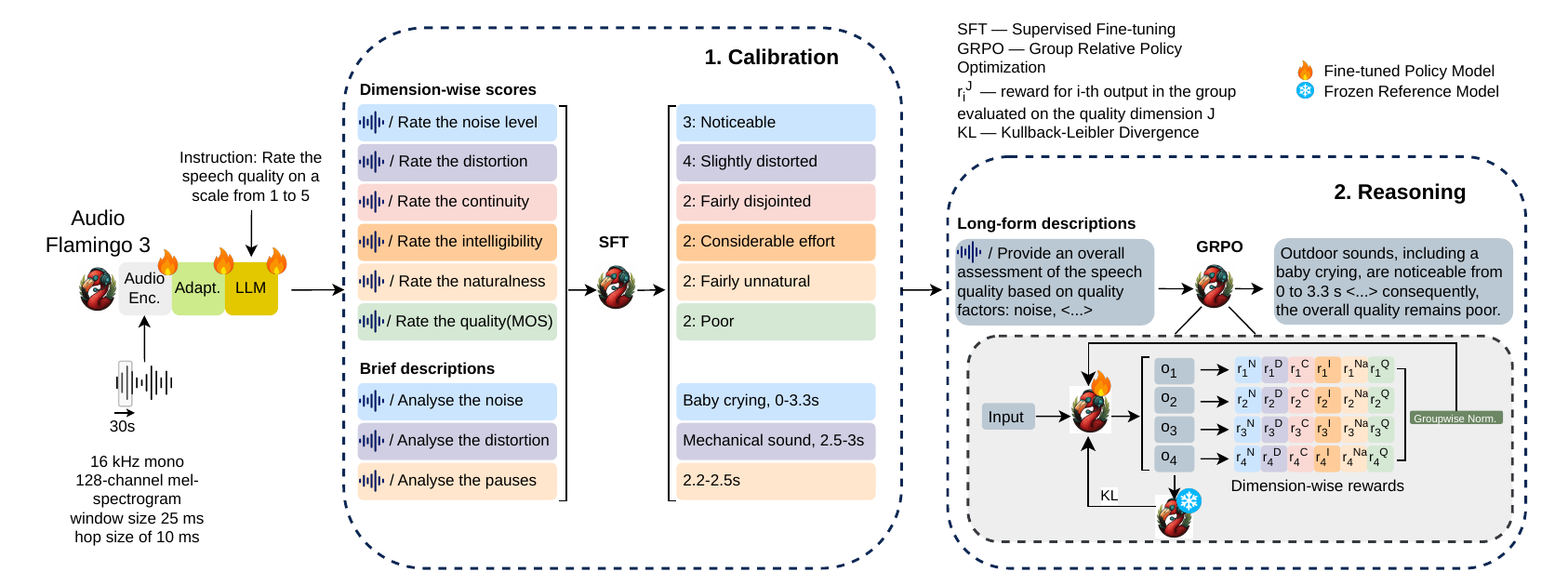}
    \caption{Overview of end-to-end fine-tuning pipeline.}
    \label{end-to-end-pipeline}
\end{figure*}

\section{Related Work}
Current approaches to descriptive speech quality assessment generally rely on coupling a pretrained audio encoder for feature extraction with an LLM for text generation, followed by fine-tuning on dimension-specific annotations. Authors of QualiSpeech \cite{wang-etal-2025-qualispeech} describe a two-stage pipeline (hereafter, QualiSpeech-FT) with Supervised Fine-Tuning (SFT) in both stages, learning dimension-wise scores followed by long-form descriptions. While the classification and detection of audio artifacts improve after the second stage, dimension-wise score prediction slightly degrades. The authors cite the limited reasoning capacity of the Vicuna \cite{zheng2023judging} LLM backbone as a potential bottleneck in their multimodal system, as it is outperformed by GPT-4o-mini \cite{openai2024gpt4omini} in deriving MOS from dimension-wise scores. Another method, SpeechQualityLLM \cite{monjur2025speechqualityllmllmbasedmultimodalassessment}, jointly fine-tunes audio encoder with LLM decoder on NISQA \cite{mittag2021nisqa} dataset to predict MOS and dimension-wise scores with justification in natural language. It focuses primarily on evaluations of scores, considering generated descriptions as supplementary and leaving the quality of the textual reasoning unassessed.

To improve reasoning relative to SFT, more recent approaches integrate Reinforcement Learning (RL) into the training pipeline. The ALLD framework \cite{chen2025audio} utilizes Direct Preference Optimization (DPO)-style \cite{rafailov2023direct} reward, while also increasing the alignment between Audio LLM and expert text-only LLM outputs via token-level distillation. SQ-LLM \cite{wang2025speechllmasjudgesgeneralinterpretablespeech} applies RL after SFT on long-form descriptions, using GRPO to optimize broad, response-level rewards (e.g., \textit{Helpfulness, Accuracy}). While standard for text reasoning, this unified reward structure does not explicitly isolate feedback for individual speech quality dimensions.

To address these limitations, our proposed method explicitly optimizes the model using reward signals tied directly to individual quality dimensions. Furthermore, to prevent the reasoning bottlenecks observed in prior work, we build our framework upon a state-of-the-art Audio LLM in the 7–8 billion parameter range.

\section{Method}
Our approach is illustrated in Figure \ref{end-to-end-pipeline}. It consists of two stages: Calibration and Reasoning, which we describe below.
\subsection{Calibration}
 First, we fine-tune model for muldimensional judgments using supervised fine-tuning. For each aspect, model learns quality scores on a scale [1, 5], where higher score corresponds to the higher quality, provided with criteria for each score in the prompt. It optimizes the likelihood of ground truth answers with Cross-Entropy loss:
\begin{equation}
    \mathcal{L}^{SFT}(\theta) = - \sum_{t=1}^{T} log p_{\theta}(y_t|y_{<t}, Emb[S, x_{prompt}]))
\end{equation}
where $y_t, t \in [1, T]$ are answer tokens, $Emb[S, x_{prompt}]$ are bi-modal embeddings of speech and instruction.

The goal of this phase is twofold: to inject quality-discriminative features into the audio encoder's latent space, and to equip the LLM to interpret these signals with high fidelity. While prior work \cite{wang-etal-2025-qualispeech, wang2025speechllmasjudgesgeneralinterpretablespeech} freezes audio encoder during post-training, the core part of our calibration stage is making audio encoder trainable to increase its sensitivity to low-level speech features.

\subsection{Reasoning}
Second, we  fine-tune the calibration checkpoint to aggregate dimension-wise predictions and reason about the overall quality using GRPO algorithm. Its work principle is to collect data, update the policy, and then immediately uses that updated policy to collect better data. In the context of text generation, policy refers to output distribution of the model over vocabulary tokens $p_{\theta}$, data refers to the batch of multimodal inputs - speech recordings with instructions. For a given input $x$, the model samples a group of $G$ candidate responses, each of which is scored by a customized reward mechanism. The comparison between rewards within a group is used as training signal:
\begin{equation}
\hat{A}_{i} = \frac{R(y_i) - \bar{R}_{\text{group}}}{\sigma_{\text{group}}}
\end{equation}
where sign and magnitude of advantages $\hat{A}_{i}$ for $i$-th response serve as the measure of its relative quality. The model weights are then updated to increase probability of above-average outputs and decrease the probability of below-average ones. The loss is regularized by a Kullback-Leibler divergence penalty between output distributions of policy model $p_{\theta}$ and the frozen reference model $p_{ref}$ to prevent reward hacking:
\begin{align}
   \mathcal{L}_{GRPO}(\theta) &= - \mathbb{E}_{i} \left[ \log p_{\theta}(y_i \mid x) \hat{A}_{i} \right] \nonumber \\
   &\quad + \lambda \text{KL}(p_{\text{ref}} \parallel p_{\theta}) 
\end{align}

\subsubsection{Dimension-wise rewards}

GRPO's effectiveness relies on bounded reward functions that evaluate candidate relevance. We design our rewards around verifiable anchors extracted from the generated text: numerical quality scores and brief artifact descriptions.

First approach relies on structured quality descriptions: we extract dimension-wise scores to estimate \textbf{Accuracy} reward:
\begin{equation}
r_i^{\text{Score}} = \mathbbm{1}[s_{\text{pred}} = s_{\text{ref}}]
\end{equation}

For audio artifacts annotated with brief descriptions we calculate \textbf{Semantic Similarity} between $y_{\text{pred}}$ and the reference $y_{\text{ref}}$ by 
computing the cosine similarity between embeddings $\mathbf{e}$ generated by the sentence transformer model and mapped linearly to $[0, 1]$:
\begin{equation}
r_i^{\text{Semantic}} = \frac{1}{2} \left( \frac{\mathbf{e}_{\text{pred}} \cdot \mathbf{e}_{\text{ref}}}{\lVert \mathbf{e}_{\text{pred}} \rVert \lVert \mathbf{e}_{\text{ref}} \rVert} + 1 \right)
\end{equation}

By combining these, we balance strict factual precision with broader contextual meaning. The total reward $R$ is the sum of accuracy and semantic reward across all dimensions.

\begin{figure}[htbp]
    \centering
    \includegraphics[width=\linewidth]{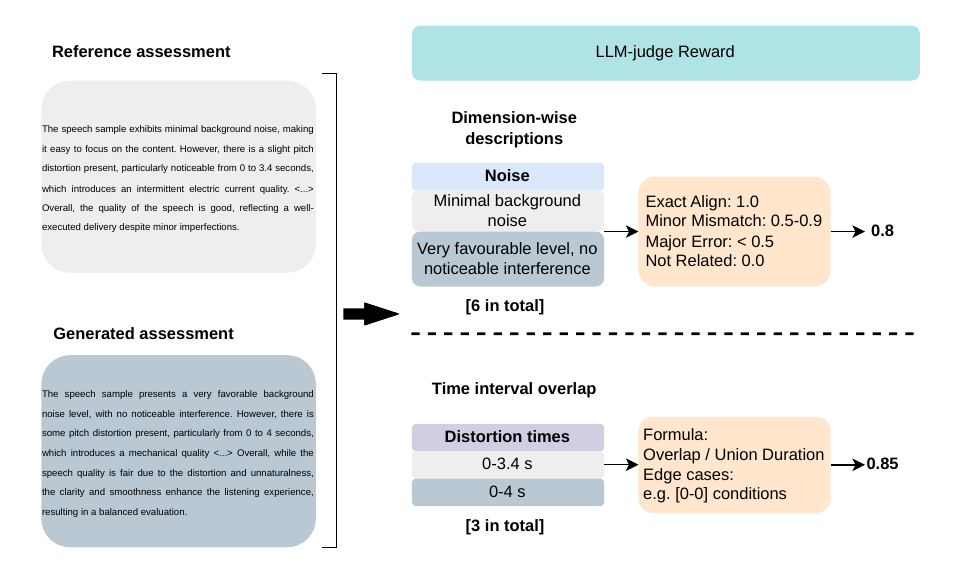}
    \caption{Dimension-wise LLM-judge reward.}
    \label{fig:judge-prompt}
\end{figure}

Second approach is to use a separate text-only LLM for multidimensional scoring. We prompt it to evaluate each generation as shown on the Figure \ref{fig:judge-prompt}. We estimate the total reward $R$ as the sum of dimension-wise rewards.

\begin{table*}[t]
\centering
\caption{Pearson Correlation Coefficients (PCC) for dimension-wise speech quality assessment across various reasoning strategies. The 'Nat.' to 'MOS' columns represent specific perceptual dimensions from the QualiSpeech dataset, 'Avg.' refers to the mean of the dimension-wise PCC values. }
\label{tab:score_assessment}
\resizebox{\textwidth}{!}{
\begin{tabular}{lccccccccc}
\toprule
\textbf{Reasoning method} & \textbf{Learned Enc.} & \textbf{Calibration} & \textbf{Nat.} & \textbf{Noise} & \textbf{Distort.} & \textbf{Effort} & \textbf{Cont.} & \textbf{MOS} & \textbf{Avg.} \\
\midrule
\multicolumn{10}{l}{\textit{Baselines}} \\
 QualiSpeech-FT-SALMONN & x & \checkmark & 0.56 & 0.65 & 0.57 & 0.61 & 0.45 & 0.63 & 0.57 \\
QualiSpeech-FT & x & \checkmark & 0.61 & 0.68 & 0.58 & 0.60 & 0.49 & 0.64 & 0.60 \\
SQ-LLM & \checkmark & \checkmark & 0.71 & 0.72 & 0.60 & 0.65 & 0.44 & 0.68 & 0.63 \\
\midrule
\multicolumn{10}{l}{\textit{Ours}} \\
Acc.+Sem. dim.-wise & \checkmark & \checkmark & 0.70 & \textbf{0.78} & 0.59 & \textbf{0.69} & 0.55 & 0.72 & 0.68 \\
LLM-judge dim.-wise & \checkmark & \checkmark & \textbf{0.73} & 0.77 & 0.68 & 0.67 & \textbf{0.63} & \textbf{0.76} & \textbf{0.71} \\
\hline
\multicolumn{10}{l}{\textit{Ablations}} \\
QualiSpeech-FT & \checkmark & \checkmark & 0.68 & 0.75 & 0.61 & 0.66 & 0.59 & 0.67 & 0.69 \\
QualiSpeech-FT & \checkmark & x & 0.55 & 0.48 & 0.59 & 0.50 & 0.35 & 0.60 & 0.51 \\
 x (Calibration-only) & \checkmark & \checkmark & 0.67 & 0.75 & \textbf{0.69} & 0.61 & 0.55 & 0.71 & 0.66 \\
\bottomrule
\end{tabular}
}
\end{table*}

\section{Experimental setup}
\subsection{Data}
We use QualiSpeech corpus for training and evaluation, as it contains both quantitative and qualitative characteristics for core perceptual dimensions. It spans 12,450 speech recordings with detailed annotations (85\% / 15\% train/test split). Listeners ratings on a scale [1, 5] are provided for seven perceptual dimensions: \textit{speed\footnote{Predictions for this aspect are intentionally excluded from the evaluation due to highly imbalanced data}, noise, distortion, naturalness, listening effort, continuity, overall quality (MOS)}. Time intervals and brief descriptions are provided for audio artifacts: \textit{noise, distortions, unnatural pauses}. Finally, all quality factors are summarized in a long natural language description.
\subsection{Models}
We adopt Audio Flamingo 3 model because of its state-of-the-art performance on advanced audio perception and reasoning benchmark MMAU \cite{sakshi2024mmaumassivemultitaskaudio} among open-weight models of similar size. We also fine-tune SALMONN-7B \cite{tang2024salmonn} model to reproduce the results from QualiSpeech paper.

\begin{table}[htbp]
\centering
\caption{Hyperparameter configurations across different fine-tuning stages. 'Gr.s.', 'B.s.', 'L.r.' refer to group size in GRPO, batch size and learning rate.}
\label{tab:hyperparameters}
\small
\begin{tabular}{llccc}
\hline
\textbf{FT Stage} & \textbf{N iter.} & \textbf{Gr. s.} & \textbf{B.s.} & \textbf{L.r.} \\
\hline
Calibration & 10k & -- & 8 & 1.5e-5 \\
SFT Reasoning & 10k & -- & 8 & 1.5e-5 \\
GRPO Reasoning & 200 & 4 & 4 & 5e-6 \\
\hline
\end{tabular}
\end{table}

\subsection{Baselines and Alignment Methodologies}
We frame the explanation of MOS as a reasoning task and categorize alignment methodologies based on their approach to reasoning: SFT or RL. We select QualiSpeech-FT and SQ-LLM as baselines because their fine-tuning curricula are comparable to our Calibration-Reasoning framework.

In the SFT-based methodologies, the second (Reasoning) stage is chain-of-thought fine-tuning on long-form descriptions. We reproduce QualiSpeech-FT with AF3 base model to evaluate the impact of more capable LLM backbone. We additionally compare frozen and end-to-end learned audio encoders and single-stage pipelines as an ablation study.

\begin{table*}[t]
\centering
\caption{Descriptive accuracy of various reasoning strategies. We report F1 score for artifact detection, Correlation score (Corr.) estimated by GPT-4o for artifact descriptions, and Intersection over Union (IoU) for overlap between detected and reference time intervals; ROUGE-L and Correlation score for long-form descriptions.}
\label{tab:desc-assessment}
\begin{tabular}{l ccc ccc ccc cc} 
\toprule
\multirow{3}{*}{\textbf{Reasoning method}} & \multicolumn{9}{c}{\textbf{Brief desc.}} & \multicolumn{2}{c}{\textbf{Long-form desc.}} \\
\cmidrule(lr){2-10} \cmidrule(lr){11-12}
 & \multicolumn{3}{c}{\textbf{Noise}} & \multicolumn{3}{c}{\textbf{Distortion}} & \multicolumn{3}{c}{\textbf{Unnatural Pause}} & \multirow{2}{*}{\textbf{ROUGE-L}} & \multirow{2}{*}{\textbf{Corr.}} \\
 \cmidrule(lr){2-4} \cmidrule(lr){5-7} \cmidrule(lr){8-10} 
 & \textbf{F1} & \textbf{Corr.} & \textbf{IoU} & \textbf{F1} & \textbf{Corr.} & \textbf{IoU} & \textbf{F1} & \textbf{Corr.} & \textbf{IoU} & & \\
\midrule
\multicolumn{12}{l}{\textit{Baselines}} \\
QualiSpeech-FT-SALMONN & 0.44 & 0.49 & 0.49 & 0.49 & 0.67 & 0.78 & 0.53 & 0.40 & 0.34 & 0.38 & 0.75 \\
QualiSpeech-FT & 0.56 & 0.52 & 0.44 & 0.73 & 0.71 & 0.81 & 0.56 & 0.40 & 0.25 & 0.40 & 0.78 \\
SQ-LLM & 0.72 & 0.57 & 0.56 & 0.82 & 0.68 & 0.74 & 0.61 & 0.39 & 0.33 & 0.45 & 0.82 \\
\midrule
\multicolumn{12}{l}{\textit{Ours}} \\
Acc.+Sem. dim.-wise & \textbf{0.77} & 0.68 & 0.61 & 0.63 & 0.70 & 0.80 & 0.54 & 0.42 & 0.46 & 0.47 & 0.80 \\
LLM-judge dim.-wise & \textbf{0.77} & 0.75 & \textbf{0.72} & 0.84 & 0.85 & \textbf{0.84} & \textbf{0.67} & \textbf{0.44} & \textbf{0.47} & \textbf{0.51} & \textbf{0.83} \\
\midrule
\multicolumn{12}{l}{\textit{Ablation}} \\
x (Calibration-only) & 0.75 & \textbf{0.85} & 0.67 & \textbf{0.88} & \textbf{0.90} & 0.81 & 0.56 & \textbf{0.44} & 0.45 & 0.12 & 0.19 \\
\bottomrule
\end{tabular}
\end{table*}

For our RL-based approaches, we apply GRPO after a single epoch of supervised fine-tuning (SFT) warm-up on long-form descriptions to stabilize the model's adherence to the expected response structure. We distinguish our fine-grained pipeline from the SQ-LLM baseline by shifting from a \textbf{unified} preference strategy, which evaluates broad properties such as \textit{"Helpfulness," "Relevance," "Detail," and "Accuracy"} across all dimensions simultaneously, to a \textbf{dimension-wise} strategy. In our approach, rewards are explicitly mapped to individual QualiSpeech quality dimensions. Following the SQ-LLM setup, we utilize Qwen3 \cite{yang2025qwen3technicalreport} as the \textbf{LLM-judge}. Furthermore, to investigate the role of dimension-specific optimization, we implement an alternative GRPO configuration driven by \textbf{Accuracy} and \textbf{Semantic Similarity} rewards. These rewards are computed from the parsed components of structured responses, providing a more interpretable and computationally efficient alternative to the LLM-judge. Semantic similarity embeddings are generated using the all-MiniLM-L6-v2 model \cite{reimers-2019-sentence-bert}.

We summarize the training configurations in Table \ref{tab:hyperparameters}. All models are trained using early stopping, with Low-Rank Adaptation \cite{hu2022lora} applied to all linear layers at a rank of $r=64$.

\section{Results}

To evaluate quality of explanations, we adopt testing approach from QualiSpeech paper: we prompt GPT-4o \cite{hurst2024gpt4o} to extract verifyable signals, such as scores, brief descriptions and time intervals from the generated text.

Table \ref{tab:score_assessment} presents the dimension-wise evaluation scores, placing special emphasis on MOS prediction, as it is derived from the quality dimensions. Our LLM-judge with dimension-wise rewards method demonstrates the strongest overall performance, surpassing previous state-of-the-art methods to achieve the highest average Pearson Correlation Coefficient (PCC) of 0.71 and a MOS PCC of 0.76. Among the RL-based approaches, both of our dimension-wise reward strategies strictly outperform the unified reward baseline (SQ-LLM) across MOS and average scores. To evaluate the impact of specific method components, we conducted a series of ablation studies. First, examining the SFT baselines reveals that upgrading the LLM backbone from SALMONN to AF3 (QualiSpeech-FT) yielded a marginal 0.03 gain in average PCC, whereas unfreezing the audio encoder provided a substantial 0.12 boost. This indicates that the task-specific alignment of acoustic features is significantly more beneficial for accurate MOS derivation than merely scaling the language model. Finally, single-stage ablations shows that a Reasoning-only model suffers a significant degradation of up to 0.20 PCC in dimension-wise predictions. While a Calibration-only model maintains a strong average PCC of 0.66, it is structurally incapable of producing natural language justifications, as we show later.

Table \ref{tab:desc-assessment} presents the descriptive accuracy metrics for characterizing audio artifacts. Brief descriptions are evaluated via F1 scores, GPT-4o estimated correlation, and Intersection over Union (IoU) for temporal localization across noise, distortion, and unnatural pauses, while long-form descriptions are assessed using ROUGE-L \cite{lin-2004-rouge} and GPT-4o correlation. Compared to existing baselines, our dimension-wise reward approaches demonstrate clear superiority. The LLM-judge dimension-wise model is the strongest overall, achieving the highest long-form ROUGE-L and correlation scores, as well as the most precise IoU for noise and unnatural pauses. Additionally, our Accuracy and Semantic Similarity dimension-wise reward approach serves as a strong second-best, outperforming both the SFT-based methods and the unified reward in SQ-LLM. Ablation studies highlight the necessity of the two-stage pipeline: while the Calibration-only model achieves high numerical accuracy on brief descriptions, it collapses on long-form generations due to its structural constraint to short answers. Most importantly, our dimension-wise approaches maintain the numerical precision gained during Calibration, successfully avoiding the degradation observed in both QualiSpeech and SQ-LLM. Overall, these findings indicate that dimension-specific rewards prevent the model from conflating distinct audio artifacts, yielding better descriptive quality and time localization accuracy.

While the proposed Calibration-Reasoning framework significantly improves diagnostic precision, it presents certain limitations. First, unfreezing the audio encoder during the Calibration stage introduces a substantial computational overhead compared to methods that utilize frozen encoders. Second, our fine-grained reward design intrinsically depends on the predefined artifact taxonomy of the QualiSpeech benchmark; consequently, the model may struggle to accurately reason about novel, out-of-distribution degradations (e.g., ultra low-bitrate codec artifacts) not present in the training distribution. 

\section{Conclusion}

In conclusion, our study highlights the critical architectural and training components required for interpretable speech quality assessment. We demonstrate that a two-stage Calibration-Reasoning pipeline, utilizing an end-to-end trainable Audio LLM, effectively balances numerical precision with rich descriptive capabilities. Making GRPO rewards dimension-aware proved pivotal: this fine-grained approach achieved a MOS PCC of 0.76 on QualiSpeech benchmark, marking a 13\% improvement over the previous SFT SOTA, while simultaneously yielding the highest accuracy of dimension-wise predictions and long-form description agreement. Furthermore, while an LLM-judge provided the most reliable training signals for GRPO, a simpler Accuracy and Semantic Similarity reward also proved highly competitive. By preserving calibration accuracy while unlocking structured reasoning, our dimension-wise reward framework proves highly effective for localized diagnostics.

In future work, we plan to extend this framework to broader acoustic domains, including music and spatial audio. Additionally, we aim to augment the GRPO reasoning stage with programmatic, signal-processing-based rewards (e.g., algorithmic clipping or dropout detectors) to further ground the LLM's temporal predictions in objective acoustic reality, reducing the reliance on computationally expensive LLM-judge reward pipelines.

\section{Generative AI Use Disclosure}
The authors acknowledge the use of Google Gemini model to polish the writing (e.g. find grammar issues) and assist with the LaTeX formatting of tables and figures.

\bibliographystyle{IEEEtran}
\bibliography{mybib}

\end{document}